\documentclass[letterpaper]{article}
\usepackage{jheppub}
\pdfoutput=1

\usepackage{amsmath}
\usepackage{amsfonts}
\usepackage{amsthm}
\usepackage{amssymb}
\usepackage{latexsym, array,multirow,  verbatim, enumerate}
\usepackage{slashed}
\usepackage{cancel}
\usepackage{dcolumn}
\usepackage{bm}
\usepackage{graphicx, caption, subcaption}  
\usepackage{url}
\usepackage[normalem]{ulem}
\usepackage{enumitem}
\usepackage{bbding}
\usepackage{tikz}
\usepackage{pifont}

\def\nn{\nonumber}

\DeclareMathOperator{\csch}{csch}

\newcommand{\ef}{\text{\tiny{E}}}
\newcommand{\mf}{\text{\tiny{B}}}
\newcommand{\Mmcp}{m_\chi}
\newcommand{\spp}{_{\text{\tiny{SPP}}}}
\newcommand{\ns}{_{\text{\tiny{NS}}}}
\newcommand{\mg}{_{\text{\tiny{M}}}}
\newcommand{\pg}{_{\text{\tiny{pol.}}}}
\newcommand{\ohm}{_{\text{\tiny{ohm}}}}
\newcommand{\hall}{_{\text{\tiny{hall}}}}

\begin{document}

\title{Novel Astrophysical Probes of Light Millicharged Fermions through Schwinger Pair Production}

\author[a,b]{Mrunal Korwar}
\author[a]{and Arun M. Thalapillil}

\affiliation[a]{Indian Institute of Science Education and Research, Homi Bhabha road, Pashan, Pune 411008, India.}
\affiliation[b]{Department of Physics, University of Wisconsin-Madison, Madison, WI 53706, USA.}
\emailAdd{mkorwar@wisc.edu}
\emailAdd{thalapillil@iiserpune.ac.in}

\date{\today}

\abstract{
The extreme properties of neutron stars provide unique opportunities to put constraints on new particles and interactions. In this paper, we point out a few interesting ideas that place constraints on light millicharged fermions, with masses below around an eV, from neutron star astrophysics. The model-independent bounds are obtained leveraging the fact that light millicharged fermions may be pair produced copiously via non-perturbative processes in the extreme electromagnetic environments of a neutron star, like a Magnetar. The limits are derived based on the requirement that conventional Magnetar physics not be catastrophically affected by this non-perturbative production. It will be seen that Magnetar energetics, magnetic field evolution and spin-down rates may all be influenced to various degrees by the presence of the millicharged particles.
}
\maketitle

\section{Introduction}
The Standard Model of particle physics has been an incredibly successful theory whose predictions have been tested to an immaculate degree. Nevertheless, it is considered incomplete and one of the foremost hints in this direction is the presence of dark matter in the universe.

An interesting possibility for a dark matter component are milli-charged particles (mCPs) -- particles carrying fractional electric charges\,\cite{Goldberg:1986nk,Cheung:2007ut,Feldman:2007wj}. They arise naturally in a large class of Standard Model extensions \,\cite{Holdom:1985ag, Dienes:1996zr, Abel:2003ue,Batell:2005wa,Aldazabal:2000sa,Abel:2008ai} and have been subjects of intense investigations in the context of observational anomalies in the recent past\,\cite{Zavattini:2005tm,Adriani:2008zr,Chang:2008aa}. They are also intriguing from the viewpoint of charge quantisation. Recently, mCPs in the mass range of a few GeV have also garnered attention due to the anomalous 21-cm absorption profile observed by the EDGES collaboration\, \cite{Bowman:2018yin}, and its possible theoretical interpretation in terms of mCPs\,\cite{Barkana:2018lgd}. All these reasons make mCPs of much current interest\,\cite{Jaeckel:2010ni,Collar:2012olx}.

In this paper, we consider the effects of light mCPs ($m_{\text{\tiny{mCP}}}\lesssim1\,\mathrm{eV}$) on neutron stars, via their non-perturbative production. The key idea is that the light mCP states could be Schwinger pair produced in the neutron star regions with large electric fields and distort the energetics of the overall system. During the completion of this work, two studies appeared placing interesting constraints on magnetic monopoles\,\cite{Hook:2017vyc,Gould:2017zwi}, through their Schwinger pair production in neutron stars. Few of their arguments are in the same spirit as ours, but as we shall see, the case for electrically charged mCPs and appropriate astrophysical considerations are very different from the monopole scenario. It will be demonstrated that the bounds obtained are relatively model independent, and more robust than stellar cooling bounds which may be evaded in certain models. mCP SPP was first considered in accelerator cavities, obtaining projected bounds of $\epsilon \lesssim 10^{-7}$ \,\cite{Gies:2006hv}. As mentioned, we will estimate limits based on the condition that non-perturbative mCP pair production does not drastically alter standard Magnetar physics. Using the constraints from neutron star energetics and related ideas, we shall demonstrate that robust, model-independent bounds as strong as $\epsilon \lesssim 10^{-12}$ may be obtained.

In Sec.\,\ref{sec:mcpconst} we briefly review the theoretical underpinnings behind mCPs and survey constraints in the low-mass region, along with caveats to these. Then, in Sec.\,\ref{sec:nsmg} we outline basic features of neutron stars and magnetars relevant for the study. In Sec.\,\ref{sec:mcpspp} we argue how non-perturbative production of mCPs in the electromagnetic environment of neutron stars could very generically affect their energetics, magnetic field evolution, and spin-down rates, thereby placing strong non-trivial bounds on mCPs. We summarise our results and conclude in Sec.\,\ref{sec:summary}.

\section{Millicharged Particles and Constraints}{\label{sec:mcpconst}}
 mCPs may be incorporated into Standard Model extensions directly, or more naturally through kinetic mixing with a singlet dark sector (denoted by $D$). In latter scenarios, there may be new gauge groups in the dark sector whose gauge fields mix with the gauge fields of the Standard Model through kinetic mixing\,\cite{Holdom:1985ag}, i.e. gauge kinetic terms in the Lagrangian are off-diagonal. Fermion mCPs are particularly attractive since chiral symmetry may render their masses small in a natural way. We are interested in this low-mass region and will assume the mCPs to be fermions.
  
In the simplest case with a single $U(1)_{\text{\tiny{D}}}$ gauge field $A^{\text{\tiny{D}}}_\alpha$, that is massless, the Lagrangian density is
\begin{equation}
\mathcal{L} \supset \bar{\chi}_{\text{\tiny{D}}}\left( i\slashed{\partial}-e_{\text{\tiny{D}}} \slashed{A}^{\text{\tiny{D}}}- \Mmcp\right) \chi_{\text{\tiny{D}}}-\frac{1}{4}A^{\text{\tiny{D}}}_{\alpha\beta}A^{{\text{\tiny{D}}}\,\alpha\beta}-\frac{\xi}{2} A^{\text{\tiny{D}}}_{\alpha\beta}B^{\alpha\beta} \; .
\label{eqn:mcp_kinmix_lag}
\end{equation}
The Standard Model particles are singlets under $U(1)_{\text{\tiny{D}}}$. $\chi_{\text{\tiny{D}}}$ is a dirac fermion in the dark sector, of mass $\Mmcp$, charged under $A^{\text{\tiny{D}}}_{\alpha}$ with a coupling $e_{\text{\tiny{D}}}$. $B_{\alpha}$ is the hypercharge $U(1)_{\text{\tiny{Y}}}$ gauge field. Field strengths are defined as $X_{\alpha\beta} \equiv \partial_\alpha X_\beta - \partial_\beta X_\alpha$. 

The last term in Eq.\,\eqref{eqn:mcp_kinmix_lag} is a kinetic mixing term\,\cite{Holdom:1985ag} between $U(1)_{\text{\tiny{D}}}$ and Standard Model hypercharge $U(1)_{\text{\tiny{Y}}}$. $\xi$ is generated at some high scale via loop-diagrams involving massive particles, charged under both $U(1)_{\text\tiny{Y}}$ and $U(1)_{\text{\tiny{D}}}$. The field redefinition $A^{\text{\tiny{D}}}_\alpha \rightarrow A^{\text{\tiny{D}}}_\alpha - \xi B_\alpha$ makes the gauge kinetic term canonical and eliminates the mixing term. This now leads to an effective coupling of the $\chi_{\text{\tiny{D}}}$ fermions to $B_\alpha$ with effective charge $\xi e_{\text{\tiny{D}}}$, which could be fractional and very small\,\cite{Holdom:1985ag}. After electroweak symmetry breaking, these $\chi_{\text{\tiny{D}}}$ therefore couple to the $U(1)_{\text{\tiny{QED}}}$ photon with a small, fractional electromagnetic charge of magnitude $\xi e_{\text{\tiny{D}}} \cos\theta_{\text{\tiny{W}}}$ ($\theta_{\text{\tiny{W}}}$ is the electroweak mixing angle). This in units of electron charge ($e$) is 
\begin{equation}
\epsilon \equiv \xi \frac{e_{\text{\tiny{D}}}}{e} \cos\theta_{\text{\tiny{W}}} \; .
\end{equation}

 \begin{figure}
 \centering
\includegraphics[width=0.65\textwidth]{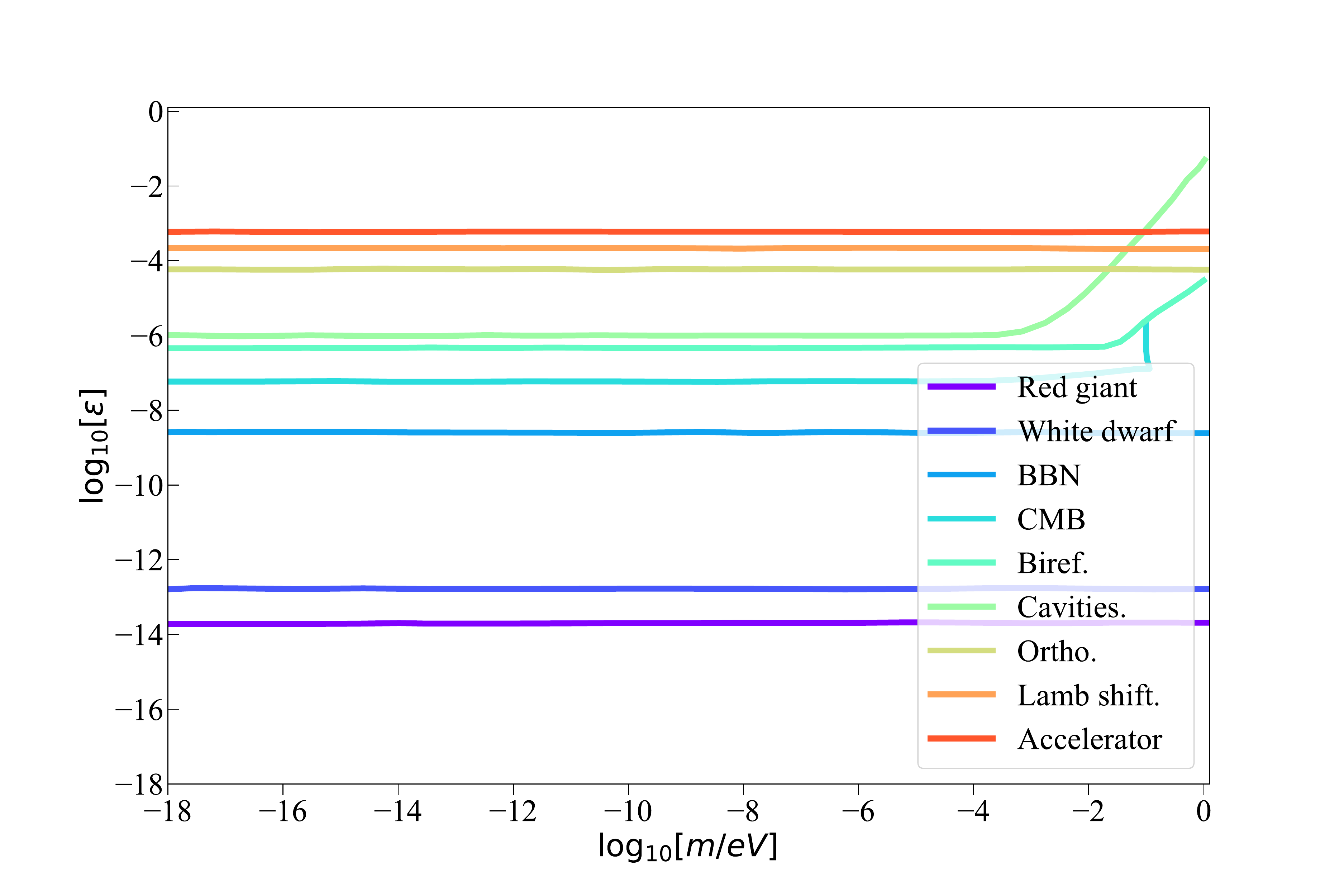}
 \caption{Constraints on light mCPs from various terrestrial, astrophysical and cosmological considerations\,\cite{Collar:2012olx, Jaeckel:2010ni,Davidson:1993sj,Davidson:2000hf, Vogel:2013raa,Foot:2014uba}. The most stringent limits are from Red giant and White dwarf stellar cooling constraints, and seemingly restrict the viable mCP charge to $\epsilon \lesssim 10^{-14}$.}
 \label{fig:constr}
 \end{figure}

The mCP parameter space $(\Mmcp,\epsilon)$ is tightly constrained by various laboratory, cosmological and astrophysical bounds\,\cite{Collar:2012olx, Jaeckel:2010ni,Davidson:1993sj,Davidson:2000hf, Vogel:2013raa,Foot:2014uba}, which are illustrated in Fig.\,\ref{fig:constr} for the relevant parameter space region. The most stringent limits are from red giant and white dwarf stellar cooling considerations. If mCPs are present, they could be produced in the stellar plasma and take away significant energy, altering conventional stellar evolution histories. One seemingly requires\,\cite{Collar:2012olx, Jaeckel:2010ni,Davidson:1993sj,Davidson:2000hf, Vogel:2013raa,Foot:2014uba}
\begin{equation}
\epsilon \lesssim 10^{-14} \; ,
\end{equation}
 to be viable, based on these stellar cooling arguments.

These astrophysical limits\,\cite{Davidson:1993sj,Davidson:2000hf, Vogel:2013raa,Foot:2014uba} nevertheless have some model dependence and may be evaded in various cases\,\cite{Masso:2006gc, Abel:2006qt, Foot:2007cq, Melchiorri:2007sq}. Many of these models have more than one $U(1)_{\text{\tiny{D}}}$ gauge group, and associated gauge field, with the feature that the effective mCP charge (say $q_{\text{\tiny{mCP}}}(k^2)$ for momentum transfer $k$) in plasma differs significantly from that in vacuum\,\cite{Masso:2006gc}
\begin{equation}
q_{\text{\tiny{mCP}}}(\omega_{\text{\tiny P}}^2) \ll q_{\text{\tiny{mCP}}}(0)\equiv \epsilon \; .
\end{equation}
Here, $\omega_{\text{\tiny P}}^2$ is the plasma frequency. This charge screening in plasma renders the stellar cooling bounds impotent, but these models have to possibly contend with some fine-tuning as well\,\cite{Masso:2006gc}. In general, viable mCP couplings all the way up to $\epsilon\sim 10^{-7}$ or larger\,\cite{Masso:2006gc, Abel:2006qt, Foot:2007cq, Melchiorri:2007sq} may, therefore, be possible. The limits we motivate will in general be immune to this charge screening scenario and hence relatively model independent. We will generically refer to fermion mCPs, in any model, as $\chi_{\text{\tiny{D}}}$.

\section{Neutron Stars and Magnetars}{\label{sec:nsmg}}
Neutron stars (NS) are supernovae collapse end products of very massive stars\,\cite{1934PNAS...20..254B,1934PNAS...20..259B}. Isolated neutron stars, not part of a binary system, may be categorised into radio pulsars and X-ray pulsars\,\cite{Becker:2009, Mereghetti:2008je}. Radio pulsars are thought to be rotationally powered (i.e. rotational energy losses power their electromagnetic emissions). The second category consists loosely of two groups -- soft-gamma repeaters and anomalous X-ray pulsars\,\cite{Mereghetti:2008je} and exhibits both persistent emissions as well as short-lived burst activities. This latter category may be accommodated in the so-called Magnetar model\,\cite{1992ApJ...392L...9D,1993ApJ...408..194T,Thompson:1995gw}. In a Magnetar (MG), the persistent luminosities and  burst activities are thought to be powered by the dissipation and decay of super-strong magnetic fields\,\cite{1992ApJ...392L...9D,1993ApJ...408..194T,Thompson:1995gw}.

\begin{figure}
 \centering
\includegraphics[width=0.7\textwidth]{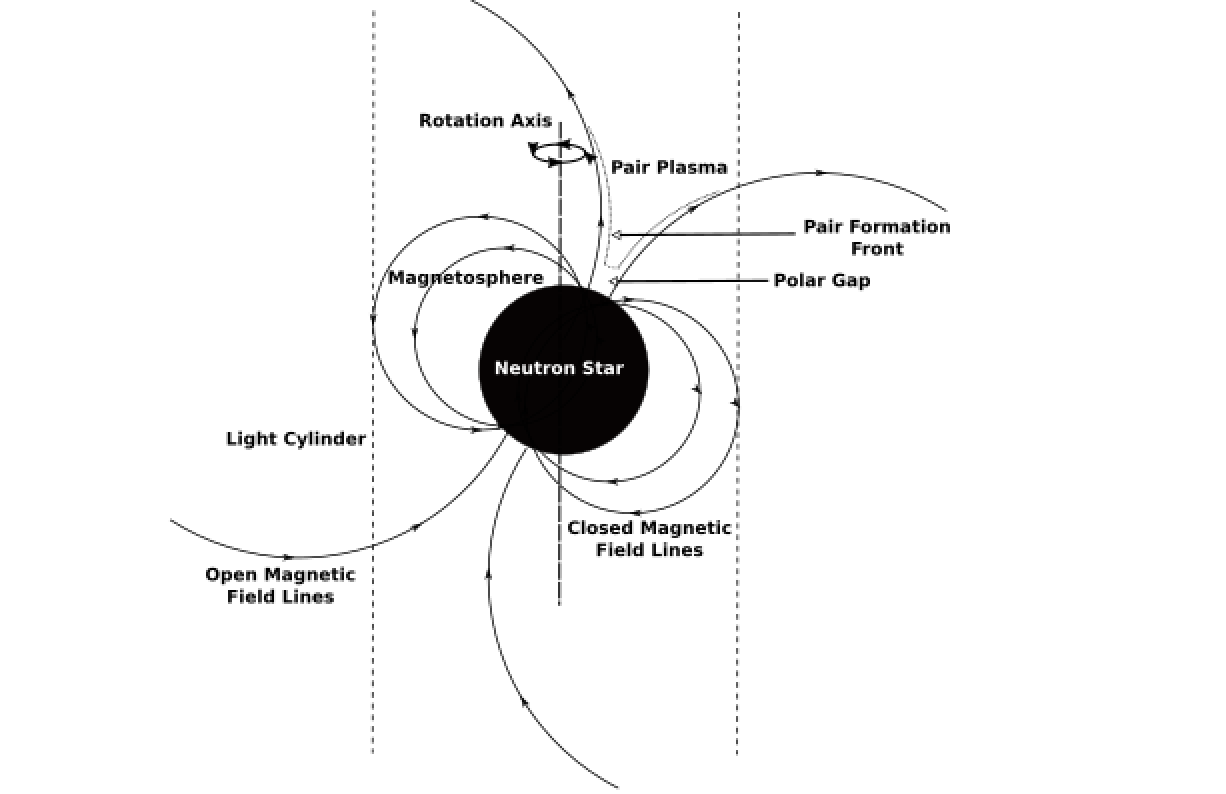}
 \caption{A schematic representation of the various relevant NS regions. Over the polar gap regions, the charge density is low and very high electric fields are generically present.}
 \label{fig:pulsar_regions} 
 \end{figure}

NS are compact, rotating objects with large magnetic fields in general. Radio pulsars already are thought to have magnetic fields typically approaching $10^{11}-10^{13}\,\mathrm{G}$, while Magnetars are thought to have even larger fields in a range $10^{14}-10^{15}\,\mathrm{G}$ or higher. The NS rotation and large magnetic fields lead to the generation of large external electric fields\,\cite{Becker:2009, 2004hpa..book.....L}. The generated Lorentz forces greatly exceed the gravitational force on the surface and lead to extraction of particles from the NS surface. The extracted particles form a co-rotating envelope around the NS called the NS magnetosphere. Once this plasma forms, a force-free condition occurs -- the distribution of charges in the magnetospheric plasma shorts-out the induced electric field, $\vec{E}_{\text{\tiny{GJ}}}+( \vec{\Omega}\ns\times \vec{r})\times \vec{B}\ns= 0$. These ideas constitute the basic Goldreich-Julian model\,\cite{Goldreich:1969sb} and describes the salient principles behind NS electrodynamics. The various regions for a typical NS are illustrated in Fig.\,\ref{fig:pulsar_regions}. 

Interestingly, the force-free state is not maintained in all magnetospheric regions though, and many models generically predict the existence of `vacuum gap' regions where the plasma density is very low or vanishing\,\cite{Becker:2009, 2004hpa..book.....L}. In these regions, the Goldreich-Julian model co-rotating condition and force-free criteria break down and electric fields are non-vanishing. This is a crucial observation for the arguments we put forward.

We will be explicitly interested in the polar gap vacuum regions (see Fig.\,\ref{fig:pulsar_regions}), where large electric fields are present. The electric field in the relevant polar gap (PG) volume is mostly parallel to the magnetic field\,\cite{Becker:2009, 2004hpa..book.....L,Harding:1998ma, Dyks:2000ee}, and has an average magnitude that may be as large as\,\cite{Becker:2009, 2004hpa..book.....L,Harding:1998ma,Dyks:2000ee}
\begin{equation}
|\vec{E}^\text{\tiny{PG}}_\text{\tiny{NS}}|=\frac{1}{2}\Omega\ns B\ns R_{\text{\tiny{NS}}}  \; .
\label{eq:EfieldNS}
\end{equation}
Here, $B\ns$ is the polar magnetic field on the NS surface and $R_{\text{\tiny{NS}}}$ is the NS radius. Taking representative MG parameter values (denoted `$\text{M}$') -- rotation period $\tau\mg=  10\,\mathrm{s}$, radius $R\mg=10\,\mathrm{Km}$, and $B\mg=10^{15}\,\mathrm{G}$, one gets in the MG case
\begin{equation}
|\vec{E}_\text{\tiny{M}}| \sim 10^{14}\,\mathrm{V}\mathrm{m}^{-1}\;.
\label{eq:epg_val}
\end{equation}

The polar gap radius is approximately given by $R\pg \simeq150\,\mathrm{m}(\tau\ns/\mathrm{s})^{-\frac{1}{2}}$\,\cite{2004hpa..book.....L}, where $\tau\ns$ is the NS rotation period. Specialising to MGs, with $\tau\mg =10\,\mathrm{s}$, one obtains $R_\text{\tiny{pol.}} \simeq 50\,\mathrm{m}$. The polar gap height and characteristic slot-gap widths are determined by the pair-formation front\,\cite{Ruderman:1975ju, Harding:1998ma, Hibschman:2001kn}. The typical pair-formation front height and slot-gap width for an MG may be taken to be $\sim10\,\mathrm{m}$\,\cite{Hibschman:2001kn}. With these dimensions and assuming $|\vec{E}_\text{\tiny{M}}|$ is significant in the slot-gap at least all the way up to a height $\mathcal{O}(2 R\ns)$, we may estimate a relevant polar gap volume ($\mathcal{V}\pg$).

\section{mCP Schwinger Pair Production in Magnetars }{\label{sec:mcpspp}}
 The Schwinger pair production (SPP) formula for $\vec{B} \shortparallel \vec{E}$\,\cite{Nikishov:1970br, Kim:2003qp,Dunne:2004nc,Ruffini:2009hg,Korwar:2018euc}, gives for the average $\chi_{\text{\tiny{D}}}$ pair production rate per unit volume,
\begin{equation}
\Gamma^{\ef \mf}_{\chi \bar{\chi}}= \frac{\epsilon^2 e^2 E B}{4\pi^2 \hbar^2} \coth\left[\frac{ \pi c B}{E}\right] \exp\left[-\frac{ \pi \Mmcp^2 c^3}{\hbar \epsilon e E}\right] \; .
\label{eq:eb_spp}
\end{equation}
The expression may be obtained in a straightforward way using worldline instanton techniques\,\cite{Korwar:2018euc}. In the MG polar gap, we have $\vec{B}\mg \shortparallel \vec{E}_\text{\tiny{M}}$ and Eq.\,\eqref{eq:eb_spp} is valid even in this strong-field regime\,\cite{Ruffini:2009hg}. Eq.\,\eqref{eq:eb_spp} goes over to the pure-$\vec{E}$ SPP result\,\cite{Schwinger:1951nm} as $B\rightarrow 0$. For mCP fermions, the additional $\vec{B}\mg$ gives a slight rate enhancement. Interestingly, scalar SPP would be suppressed for field values $\vec{E}_\text{\tiny{M}}$ and $\vec{B}\mg$-- the $\coth\left[\pi c B/E \right] $ factor in Eq.\,\eqref{eq:eb_spp} gets replaced by $\csch\left[\pi c B/E\right]$ for scalars\,\cite{Kim:2003qp,Dunne:2004nc,Ruffini:2009hg}. 

\begin{figure}
 \centering
\includegraphics[width=0.55\textwidth]{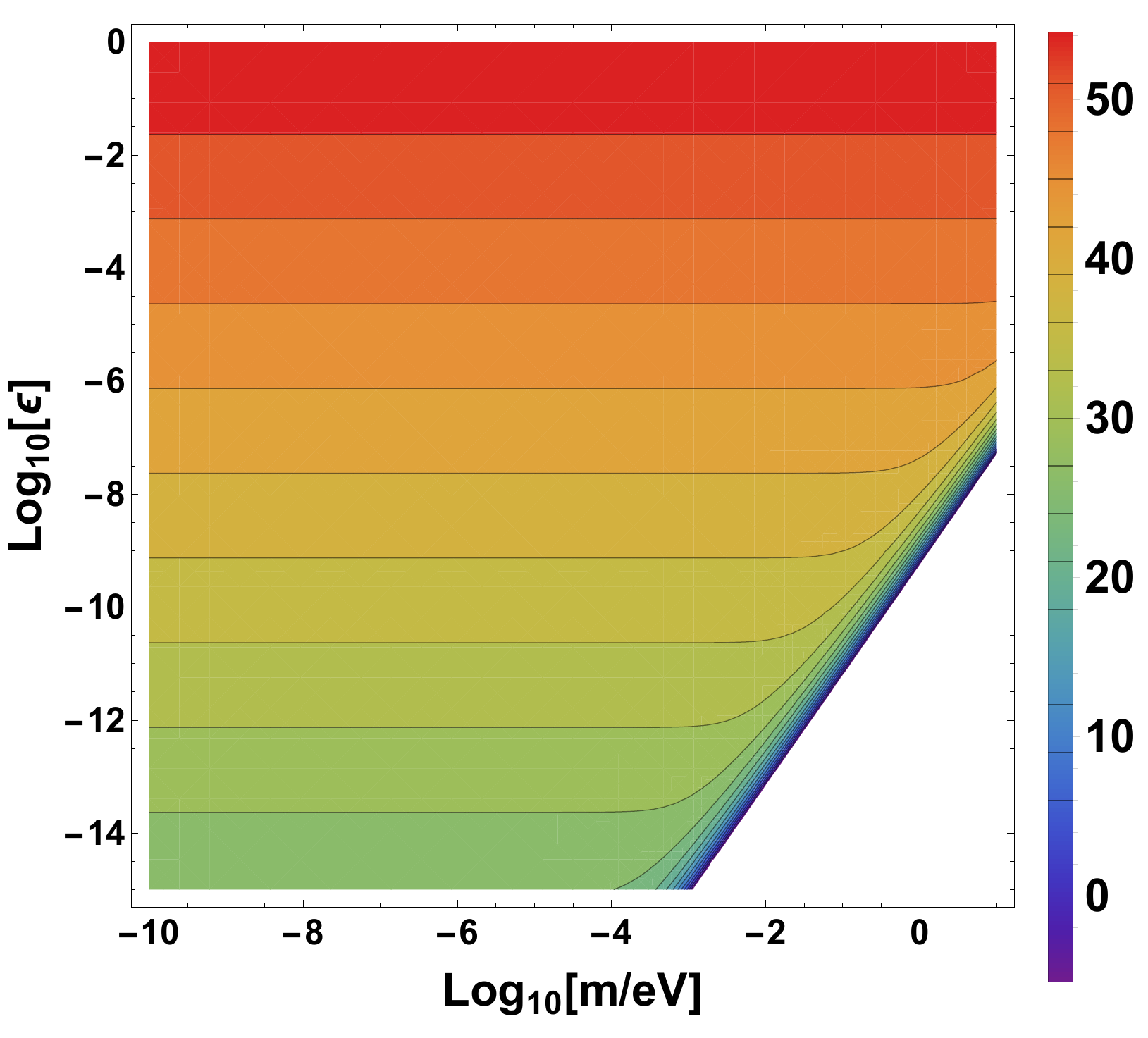}
 \caption{Pair production rates for light millicharged fermions in a representative polar gap electric field of $10^{14}\,\mathrm{V}\mathrm{m}^{-1}$. The rates per unit volume are in units of $\mathrm{m^{-3} \,s^{-1}}$. The typical rates in most of the parameter space of interest are many orders of magnitude larger than the corresponding $e^+-e^-$ rate at these field values. The latter requires fields close to a critical value of $10^{18}\,\mathrm{V}\mathrm{m}^{-1}$, for Schwinger pair production to be significant.}
 \label{fig:ebspprate}
 \end{figure}

The rates based on Eq.\,\eqref{eq:eb_spp}, for $B\mg=10^{15}\,\mathrm{G}$ and induced $|\vec{E}\text{\tiny{M}}|=10^{14}\,\mathrm{V}\mathrm{m}^{-1}$, are significant and many orders of magnitude larger than $e^+e^-$-SPP rates for these field values (see Fig.\,\ref{fig:ebspprate}). One also observes from Fig.\,\ref{fig:ebspprate} that the rates are appreciable only for masses below $\mathcal{O}(1\,\mathrm{eV})$, due to the exponential suppression. This will, therefore, be a natural boundary for our study, as mentioned earlier. 

This non-perturbative pair production has the potential to affect MG energetics. This is the central idea of the paper. Electromagnetic vacuum boundary conditions\,\cite{Goldreich:1969sb} in the polar gap\,\cite{Becker:2009, 2004hpa..book.....L,Harding:1998ma,Dyks:2000ee} broadly ensure that the shorting of the electric fields will be accompanied by persistent electromagnetic energy losses. The polar gap $E$-field must be regenerated by the magnetic field and rotation. As we shall argue later, the change in angular momentum due to mCP-SPP and subsequent evolution is very marginal, and it is the electromagnetic energy reservoir which is the main power source. Even in the hypothetical scenario where there is a catastrophic collapse of the pair-formation front (Fig.\,\ref{fig:pulsar_regions}), and loss of the polar vacuum gap, say, one generally should expect that the magneto-hydrodynamic instabilities that lead to the opening of the polar and outer vacuum gaps would still be operational, leading to their recreation. The process would then repeat, with energy tapped from the electromagnetic energy reservoir. Hence, generally, one would expect some persistent electromagnetic energy loss to be present,  if mCP-SPP is operational in a Magnetar. Also note that the precise details of mCP evolution subsequent to SPP are less important as long as no significant energy is deposited back into the magnetic field. This is true to good approximation -- mCP SPP is a dissipative process.

We will work in $\hbar=c=1$ units. In MGs, as we commented earlier, the persistent luminosities are known to be powered by super-strong magnetic field decays\,\cite{1992ApJ...392L...9D,1993ApJ...408..194T,Thompson:1995gw}. These magnetically sourced radiation losses, therefore, have to be included in any consideration of MG energetics. Now, the rate of energy loss from the MG, averaged over a lifetime $\mathcal{T}\mg$, should be bound approximately by
\begin{equation}
\int d\mathcal{V}\left[\frac{d^2\mathcal{E}_{\text{\tiny{rad.}}}}{dt\,d\mathcal{V}}+\frac{d^2\mathcal{E}^{\chi \bar{\chi}}\spp}{dt\,d\mathcal{V}} \right] \lesssim \int d\mathcal{V}\frac{1}{\mathcal{T}\mg}\left[ \frac{\vec{B}^{2}\mg}{2\mu_0}+\frac{\epsilon_0 \vec{E}^{2}_\text{\tiny{M}}}{2}\right]  .
\label{eq:eloss}
\end{equation}

$d^2\mathcal{E}_{\text{\tiny{rad.}}}/dt\,d\mathcal{V}$ represents the rate of energy loss, per unit volume, due to radiation losses. The magnetic field decays in the interior of the MG are thought to drive radiation phenomena such as quiescent X-ray emissions, X-ray/Gamma-ray flares and burst events~\cite{1992ApJ...392L...9D,1993ApJ...408..194T,Thompson:1995gw,Mereghetti:2008je}. These processes, therefore, derive their energy from the electromagnetic energy reservoir, denoted by the right-hand side of Eq.(\ref{eq:eloss}).

In the sub-eV mass range of interest, the annihilation of the mCPs are not expected to contribute significantly in this spectral range and therefore to the radiation loss term. Another possibility is that one particle in the pair accelerates over a short distance and hits the NS surface. This could contribute to collisional heating of the NS. Such a contribution, to various extent, may indeed be hidden inside the radiation loss term, which shall be directly estimated from observational data. There is nevertheless no double-counting and \,Eq.(\ref{eq:eloss}) is consistent, as the SPP term included is only for energy extracted at the instant of pair production and subsequent acceleration of the other particle in the pair outward. Note also that the energy extracted by the latter phenomena, per unit time, is much larger. As we shall also see, we will estimate the radiation loss term from available observational data\,\cite{Olausen:2013bpa} of soft, persistent quiescent X-ray emissions. These may be fit to a thermal blackbody spectrum with a power-law tail or a double blackbody spectrum (see for instance discussions in\,\cite{Mereghetti:2008je, Kaspi:2017fwg} and references therein). Curvature or synchrotron radiation due to the outward accelerating mCP, should also therefore not be significantly contributing to this component. In any case, as we shall see, dropping the radiation loss term will not quantitatively change the limits in any significant way. We nevertheless retain it to be conceptually consistent and towards the possibility that an accounting of this term may improve in future, paving the way to even better limit estimates.

$d^2\mathcal{E}^{\chi \bar{\chi}}\spp/dt\,d\mathcal{V}$ similarly quantifies energy losses, per unit volume, due to potential mCP SPP -- 
\begin{equation}
\frac{d^2\mathcal{E}^{\chi\bar{\chi}}\spp}{dt~dV~~~}= \Gamma^{\ef\mf}_{\chi \bar{\chi}} \,\epsilon e |\vec{E}_\text{\tiny{M}}| l_0 + \Gamma^{\ef\mf}_{\chi \bar{\chi}} \,  \epsilon e |\vec{E}'_\text{\tiny{M}}| (l-l_0) \; .
\label{eq:spp_work}
\end{equation}
Here, $\vec{E}_\text{\tiny{M}}$ and $\vec{E}'_\text{\tiny{M}}$ are the average electric field values over the respective distance ranges. The first term in Eq.\,\eqref{eq:spp_work} is the energy extracted per unit volume per unit time from the $\vec{E}_\text{\tiny{M}}$ field for SPP. $l_0$ is the characteristic inter-mCP distance at the instant of SPP. The second term in Eq.\,\eqref{eq:spp_work} is the subsequent work that may be done by $\vec{E}'_\text{\tiny{M}}$ in accelerating one of the $\chi_{\text{\tiny{D}}}$ particles out by a distance $l-l_0$. 

In $(\Mmcp,\epsilon)$ regions where inter-mCP dark Coulombic attraction ($F^{{\text{\tiny{Coul.}}}}_{\text{\tiny{D}}}\sim e^2_{\text{\tiny{D}}}/l_o^2$) exceeds external Lorentz force ($F_{\text{\tiny{E}}}\sim\epsilon e |\vec{E}_\text{\tiny{M}}|$), no mCPs accelerate out of the MG. The pairs would instead annihilate soon after SPP. Thus, the second term in Eq.\,\eqref{eq:spp_work} gives no contribution in these regions and the only energy extracted from the electromagnetic field is to initiate SPP. Rate computations at strong coupling\,\cite{Affleck:1981bma,Affleck:1981ag,Ritus:1998jm} suggest that mCP SPP with the dark Coulombic interaction included would give a correction $\sim \exp[e_{\text{\tiny{D}}}^2/4]$ to the exponent in Eq.\,\eqref{eq:eb_spp}, and only further enhance rates. In other regions where $F_{\text{\tiny{E}}} > F^{{\text{\tiny{Coul.}}}}_{\text{\tiny{D}}}$, energy is extracted from the electromagnetic field both for SPP and to subsequently accelerate mCPs out of the MG. $e_{\text{\tiny{D}}}$ is a free parameter in general, and for our quantitative comparisons, we will assume $e_{\text{\tiny{D}}} \sim \mathcal{O}(e)$-- keeping within perturbative limits, where the corrections due to it to the instanton then may be ignored at leading order\,\cite{Affleck:1981bma,Affleck:1981ag,Ritus:1998jm} . In models where mCPs are introduced directly, without resorting to a kinetic mixing framework, considerations due to $e_{\text{\tiny{D}}}$ may also be completely discounted. Many seminal studies have taken this path\,\cite{Gies:2006hv,Li:2013pfh}, but we retain the possibility for being as general as possible.

In Eq.\,\eqref{eq:eloss}, $\int d\mathcal{V}$ denotes integration over a relevant volume for each term. An estimate of the volume, over which Eq.\,\eqref{eq:eb_spp} is valid and mCP SPP may be significant, is given by $\mathcal{V}\pg$, as discussed earlier in Sec. \ref{sec:nsmg}. For the electromagnetic energy stored in the MG, we assume that most of it is within a characteristic distance $\sim R\ns$ of the magnetosphere.

As an estimate for the radiation loss component, we take the average of the soft, persistent quiescent X-ray emissions ($\sim 2-10\,\mathrm{KeV}$), from all currently known MG candidates\,\cite{Olausen:2013bpa}. This gives
\begin{equation}
\big\langle\int d\mathcal{V}\, d^2\mathcal{E}_{\text{\tiny{rad.}}}/dt d\mathcal{V}\big\rangle_\text{\tiny{M}}~=~ 4.3\times10^{34}\,\mathrm{ergs\, s^{-1}} \; .
\label{eq:pqxrest}
\end{equation}
While making the above estimate, contributions from other components---for instance, flares and burst events~\cite{1992ApJ...392L...9D,1993ApJ...408..194T,Thompson:1995gw,Mereghetti:2008je}---are being neglected. The above quantity is therefore slightly underestimating the radiation loss contribution. As we commented earlier, Eq. (\ref{eq:pqxrest}) is quantitatively much smaller than the electromagnetic energy budget and may in principle be dropped without drastically affecting our ensuing limit estimates.

We may now put together all the above considerations to place limits. In most regions of our parameter space, shown in Fig.\,\ref{fig:ebspprate}, the exponential factor in Eq.\,(\ref{eq:eb_spp}) is $\mathcal{O}(1)$. We may, therefore, obtain approximate limits by setting it to be a constant. Deviations from this assumption are only appreciable near the boundary of SPP suppression and beyond. Equivalently, it may be solved numerically.

In regions where $F_{\text{\tiny{E}}} > F^{{\text{\tiny{Coul.}}}}_{\text{\tiny{D}}}$ and mCP SPP occurs, one gets from Eq.\,\eqref{eq:eloss} an approximate bound
\begin{equation}
\epsilon ~ \lesssim  ~10^{-12} ~~(\text{for regions with}~F_{\text{\tiny{E}}} > F^{{\text{\tiny{Coul.}}}}_{\text{\tiny{D}}})\; .
\label{eq:elossbound}
\end{equation}
This is obtained taking typical MG values $\tau\mg=  10\,\mathrm{s}$, $R\mg=10\,\mathrm{Km}$, $\mathcal{T}\mg = 10^4\,\mathrm{yrs}$ and $B\mg=10^{15}\,\mathrm{G}$ and assuming $l=20\,\mathrm{Km}$. Note that this would also be the relevant bound if mCPs were introduced directly in the Lagrangian, without any inter-mCP dark interaction. In regions where $F_{\text{\tiny{E}}} < F^{{\text{\tiny{Coul.}}}}_{\text{\tiny{D}}}$, the only energy extracted from the electromagnetic field is to achieve SPP. This situation is tantamount to putting $l=l_0$ in Eq.\,\eqref{eq:spp_work}. The characteristic inter-mCP distance $l_0$ at the instant of SPP is given by
\begin{equation}
l_0 = \frac{2 \Mmcp}{\epsilon e |\vec{E}_\text{\tiny{M}}|} \; .
\end{equation}
This is valid for both strong coupling and large fields by energy conservation and symmetry. With these considerations, Eq.\,\eqref{eq:eloss} for $F_{\text{\tiny{E}}} < F^{{\text{\tiny{Coul.}}}}_{\text{\tiny{D}}}$ gives a bound
\begin{equation}
\epsilon^2 \left(\frac{\Mmcp}{1\, \mathrm{eV}}\right) ~ \lesssim  ~10^{-16}~~(\text{for regions with}~F_{\text{\tiny{E}}} < F^{{\text{\tiny{Coul.}}}}_{\text{\tiny{D}}})\;,
\label{eq:elossbound}
\end{equation}
in regions where SPP is unsuppressed.

In the $(\Mmcp,\epsilon)$ parameter space of interest, $l_0 \ll 10\,\mathrm{m}$ and particle separations are always within the polar gap at the time of SPP. This, along with the relevant Compton wavelengths, support the validity of Eq.\,\eqref{eq:eb_spp} and the neglect of any field inhomogeneities to leading order\,\cite{Ritus:1998jm,Dunne:2004nc}.

Note that these limits only depend on the fact that fermion mCPs have an effective coupling with the $U(1)_{\text{\tiny{QED}}}$ photon. Any model dependent charge screening mechanism in plasma\,\cite{Masso:2006gc, Foot:2007cq, Melchiorri:2007sq} that makes stellar-cooling bounds weak, is also irrelevant in the the vacuum gap regions.  

 To augment the central idea from energetics, let us now regard the related, potential effects from mCP SPP on MG magnetic field decay and spin-down rates. mCP-SPP occurring in the polar vacuum gap regions should take away energy from the electromagnetic field, without significant loss of angular momentum from the system. The latter fact may be motivated concretely by considering the aligned rotor model\,\cite{Goldreich:1969sb}, and the observation that the field's angular momentum density in the polar vacuum gap is very small, as $\vec{E}\times \vec{B}\simeq 0$. Also, since after each instance of mCP SPP, the particle and anti-particle are accelerated with equal and opposite momenta, any change in angular momenta of the system due to subsequent ejections should only be secondary to the primary effect on the electromagnetic energy reservoir. Moreover, when $F_{\text{\tiny{E}}} < F^{{\text{\tiny{Coul.}}}}_{\text{\tiny{D}}}$, there is no ejection at all, and this is strictly true. 
 
 Therefore, the non-perturbative decay of the electric field and any ejection should not be extracting any appreciable angular momentum from the system. The $B$-field and the $\Omega$-rotation power the electric field in the vacuum polar gap. A non perturbative decay of this $E$-field leading to a $\Delta E$ change in the field strength, under the situation where there is very little change in angular momenta, therefore leads to a $\Delta B$, based on Eq.\,(\ref{eq:EfieldNS}). We may hence argue that there is a new contribution to $dB/dt$ from the mCP-SPP non-perturbative phenomena, based on energy and angular momentum conservation. Note that this non-perturbative process leading to field decay is very distinct from the established scenarios in neutron stars; where perturbative and conventional electrodynamic phenomena such as production of $e^+e^-$ pairs in the vacuum gaps and other magnetospheric processes are sometimes argued to not lead to changes in the magnetic field. The relevant energetic and field decay aspects of mCP-SPP are independent of the assumptions of the aligned rotor motor, and hence should also be true even when the magnetic and rotational axes are slightly misaligned. In the scenario where the axes are misaligned, conventional magnetic dipole radiation and a corresponding braking torque due to radiation reaction will be important though and will cause a spin down of the NS\,\cite{Shapiro:1983du}. Since mCP-SPP affects the average magnetic field evolution, there is now a possibility that the spin-down ($d\Omega/dt$) will be modified. These are the related ideas that we would like to explore in the context of the main idea based on energetics, as encapsulated in Eq.\,\ref{eq:eloss}.
 
 Ohmic and Hall drift contributions\,\cite{1992ApJ...395..250G, Glampedakis:2010ec} are conventionally responsible for field attenuations in the NS interior. The details of these contributions and how magnetic field configurations evolve in an NS are not fully know, and are topics of intense study (see for instance,\cite{Turolla:2015mwa, Kaspi:2017fwg} and references therein). We may nevertheless try to capture some salient features by incorporating appropriate time-scales and terms relevant in the conventional evolution of Magnetar magnetic fields (see for instance Eqs.\,(16)-(20) in\,\cite{Aguilera:2007xk}). The new possibility now is that mCP SPP may also non-perturbatively contribute to field decays in the MG, as we have motivated. We equate the energy loss due to mCP SPP to a corresponding change in the net electromagnetic energy stored in the dipole field, which powers all Magnetar processes, as
 \begin{equation}
 \frac{d}{dt}\left(\frac{1}{12} B\mg^2 R\mg^3\right)_{\text{\tiny{mCP SPP}}} \simeq~ \epsilon e E\mg \, l\, \mathcal{V}\pg \Gamma^{\ef \mf}_{\chi \bar{\chi}} \; .
 \end{equation}
One may then phenomenologically model the overall magnetic field evolution, in the toy model of the MG system, as
\begin{eqnarray}
\frac{dB\mg(t)}{dt}&\simeq& -\frac{B\mg(t)}{\tau\ohm}-\frac{B\mg^2(t)}{B\mg(0) \tau\hall}-\frac{3 \epsilon^3 e^3 \Omega\mg^2(t)   B^2\mg(t) l}{8 \pi^2 R\mg} \mathcal{V}\pg \nn \\
&&\coth\left[\frac{2\pi }{\Omega\mg(t) R\mg}\right] \exp\left[-\frac{2\pi \Mmcp^2}{ \epsilon e \Omega\mg(t) R\mg B\mg(t)}\right] \nn \\
&=&-\rho\ohm B\mg(t)-\rho\hall \frac{B^2\mg(t)}{B\mg(0)}-\rho\spp\,l \, \Omega\mg^2(t) B^2\mg(t) ~~~ \nn \\
&& \coth\left(\frac{\hat{\rho}\spp}{\Omega\mg(t)}\right) \exp\left[-\frac{\tilde{\rho}\spp}{\Omega\mg(t)B\mg(t)}\right]\; .
\label{eq:bevolv}
\end{eqnarray} 
$B\mg(0)$ is the initial magnetic field, which we take as $10^{15}\,\mathrm{G}$. $\Omega\mg(t)$ is the MG angular velocity. $\mathcal{V}\pg$ as before is the relevant polar gap volume. For the Ohmic and Hall drift time constants, we take $\tau\ohm=10^6\,\mathrm{yrs}$ and $\tau\hall=10^4\,\mathrm{yrs}$ following typical values from literature\,\cite{1992ApJ...395..250G,Aguilera:2007xk}. Realistically, the time constants are complex functions of temperature and density, but the above values have been found to phenomenologically capture relevant behavior\,\cite{Aguilera:2007xk}. Moreover, an equation of the above form, without the SPP term, is known to semi-quantitaively reproduce\,\cite{Aguilera:2007xk} results from more detailed magneto-thermal simulations\,\cite{Aguilera:2007xk,Vigano:2013lea,2007A&A...470..303P}. Note though that the basic principle is largely independent of modelling and depends only on the fact that there is a potentially new non-perturbative dissipative contribution from mCP SPP. For typical MG parameters, the coefficients in Eq.\,\eqref{eq:bevolv} are given by $\rho\ohm=2.1\times 10^{-38}\,\mathrm{GeV}$, $\rho\hall=2.1\times 10^{-36}\,\mathrm{GeV}$, $\rho\spp= 3.1\times 10^{32}\,\epsilon^3\,\mathrm{GeV}^{-2}$, $\hat{\rho}\spp=1.24\times10^{-19}\,\mathrm{GeV}$ and $\tilde{\rho}\spp=4.1\times 10^{-37}\,\epsilon^{-1}\,(\Mmcp /1\,\mathrm{eV})^2\,\mathrm{GeV}^{3}$.
 \begin{figure}
 \centering
\includegraphics[width=0.7\textwidth]{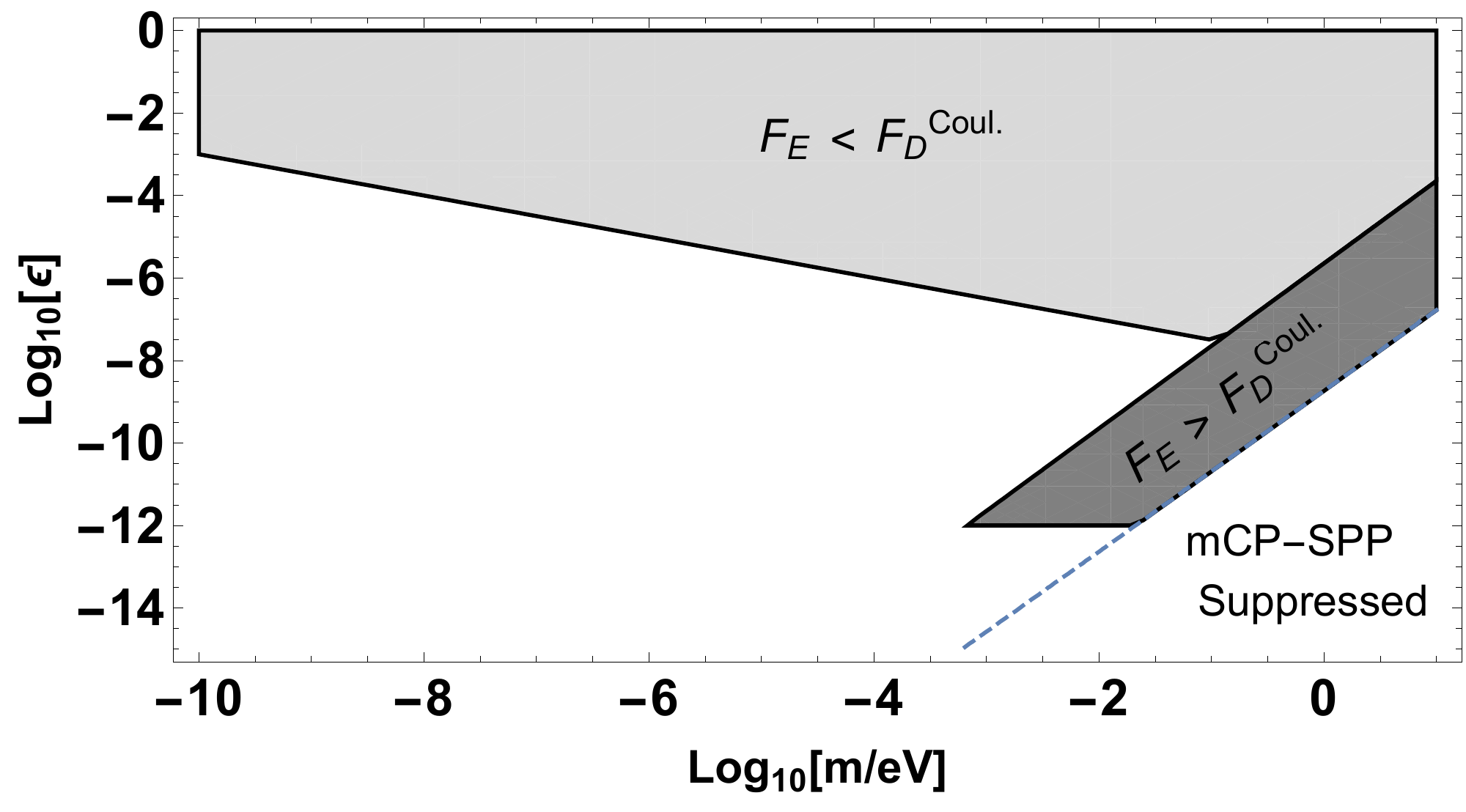}
 \caption{$\chi_{\text{\tiny{D}}}$ exclusion regions based on energy-loss and $B\mg(t)$ evolution arguments, from Eqs.\,\eqref{eq:eloss} and \eqref{eq:lossoverwhelm}. A bound $\epsilon \lesssim 10^{-12}$ is obtained for $F_{\text{\tiny{E}}} > F^{{\text{\tiny{Coul.}}}}_{\text{\tiny{D}}}$ regions, or alternatively, for all regions if mCP fermions are introduced directly without any inter-mCP dark interactions. $F_{\text{\tiny{E}}} < F^{{\text{\tiny{Coul.}}}}_{\text{\tiny{D}}}$ regions have a $m_\chi$ dependent bound on $\epsilon$. For making the comparison, we have assumed $e_{\text{\tiny{D}}} \sim \mathcal{O}(e)$.}
 \label{fig:mCP_StudyConstraints}
 \end{figure}

One reasonable supposition could be that for viable $(\Mmcp,\epsilon)$ values, the field decays due to mCP SPP should not overwhelm the conventional $B\mg(t)$ evolution in the MG. From the viewpoint of Eq.\,\eqref{eq:bevolv} a criteria could be
\begin{eqnarray}
\frac{B\mg(0)}{\tau\ohm}+\frac{B\mg(0)}{ \tau\hall} &>& \frac{3  \epsilon^3 e^3 \Omega\mg^2(0) B^2\mg(0) l}{8 \pi^2 R\mg} \coth\left[\frac{2\pi }{\Omega\mg(0) R\mg}\right]  \nn \\ 
&& \exp\left[-\frac{2\pi \Mmcp^2}{ \epsilon e \Omega\mg(0) R\mg B\mg(0)}\right]\mathcal{V}\pg \;.~~~~
\label{eq:lossoverwhelm}
\end{eqnarray}
This gives a constraint 
\begin{equation}
\epsilon~<~3.4\times10^{-12} ~~(\text{for regions with}~F_{\text{\tiny{E}}} > F^{{\text{\tiny{Coul.}}}}_{\text{\tiny{D}}})\;,
\label{eq:limlossoverwhelm1}
\end{equation}
in regions where $F_{\text{\tiny{E}}}$ dominates and SPP occurs. In regions where $F^{_{\text{\tiny{Coul.}}}}_{\text{\tiny{D}}}$ dominates, one has to set as before $l=l_0(\Mmcp,\epsilon,B\mg(0),\Omega\mg(0))$ in Eq.\,\eqref{eq:lossoverwhelm}. Therefore, in these regions a further functional dependence on the parameters ($\Mmcp,\epsilon$) and variables ($B\mg, \Omega\mg$) enters Eq.\,\eqref{eq:bevolv} through $l$. For these regions, we have
\begin{equation}
\epsilon^2 \left(\frac{\Mmcp}{1\, \mathrm{eV}}\right) ~ <  ~6.4\, \times 10^{-17} ~~(\text{for regions with}~F_{\text{\tiny{E}}} < F^{{\text{\tiny{Coul.}}}}_{\text{\tiny{D}}}) \; .
\label{eq:limlossoverwhelm2}
\end{equation}

The bounds obtained from Eqs.\,\eqref{eq:eloss} and \eqref{eq:lossoverwhelm} are comparable, which make sense -- in the conventional scenario, without mCP SPP, the Ohmic and Hall terms lead to $B\mg$ dissipation, which subsequently power persistent emissions. Thus, as suspected, our arguments based on MG energetics are related to those based on overall $B\mg(t)$ evolution in the system. The complete exclusion regions based on these arguments from Eqs.\,\eqref{eq:eloss} and \eqref{eq:lossoverwhelm} are shown in Fig.\,\ref{fig:mCP_StudyConstraints}. We have, as before, assumed for the coupling strength $e_{\text{\tiny{D}}} \sim e$.

 The change in $\Omega\ns(t)$ may be encapsulated in a `braking-index'\,\cite{Becker:2009, 2004hpa..book.....L}. For a relation $\dot{\Omega}\ns(t)\,=\,-\lambda(t)\,\Omega\ns^b(t)$, the true braking-index ($b_{\text{\tiny{true}}}$) is given by
\begin{equation}
b_{\text{\tiny{true}}}(t)=\frac{\Omega\ns(t) \ddot{\Omega}\ns(t)}{\dot{\Omega}\ns^2(t)}=b+\frac{\dot{\lambda}(t)\Omega\ns(t)}{\lambda(t) \dot{\Omega}\ns(t)} \; .
\label{eq:true_break}
\end{equation}
If $\lambda(t)$ is a constant, one obtains $b_{\text{\tiny{true}}}=b$. For instance, a rotating, constant magnetic dipole has $b_{\text{\tiny{true}}}=3$. In general, $b_{\text{\tiny{true}}}$ is time dependent as seen from Eq.\,\eqref{eq:true_break}.
 \begin{figure}
 \centering
\includegraphics[width=0.7\textwidth]{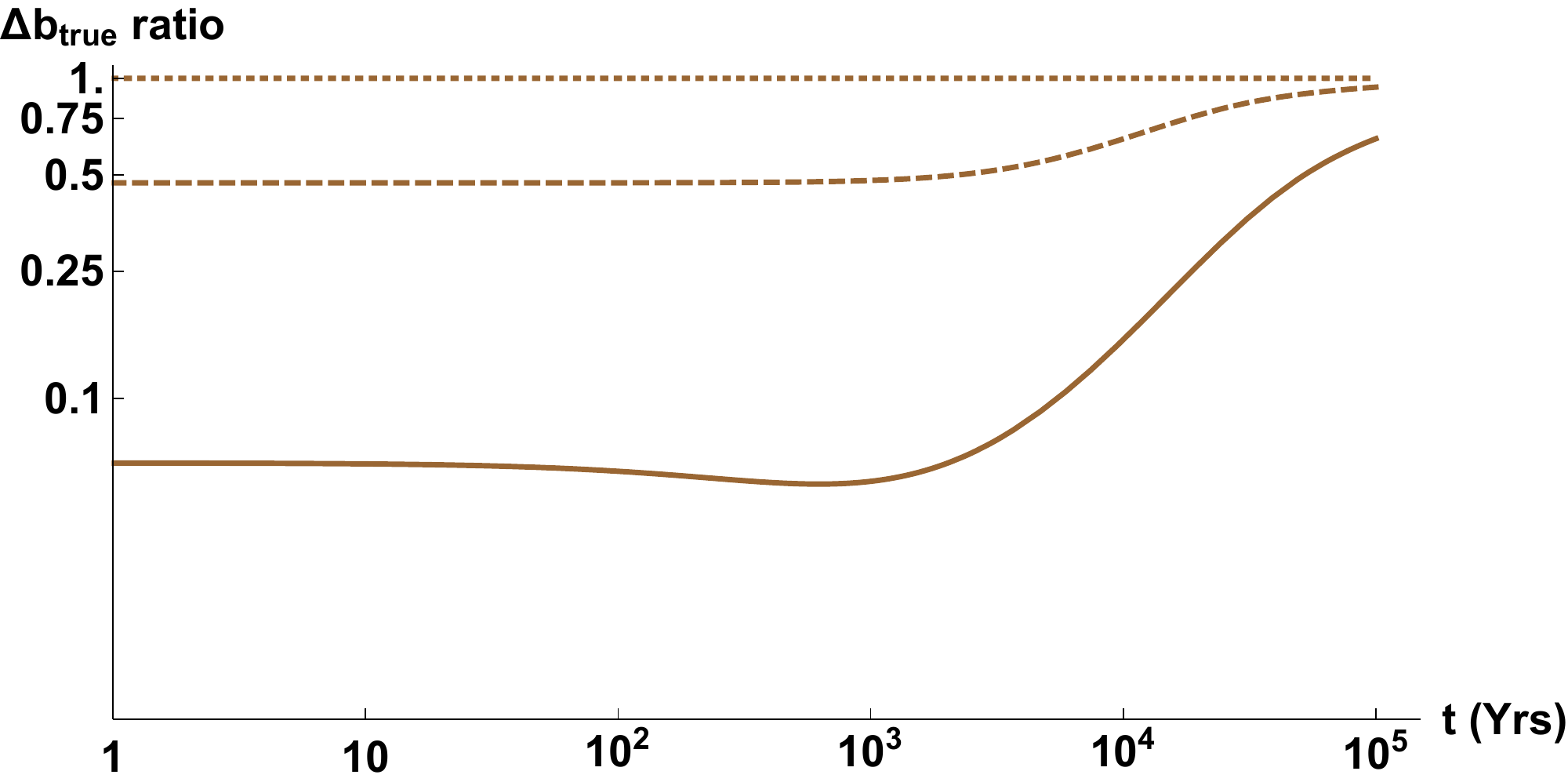}
 \caption{ The ratio  $\Delta b^{\cancel{\text{\tiny{SPP}}}}_{\text{\tiny{true}}}/\Delta b^{\text{\tiny{SPP}}}_{\text{\tiny{true}}}$ for $\Mmcp= 10^{-2}\,\mathrm{eV}$ and $\epsilon= 10^{-11}$ (solid), $10^{-12}$ (dashed) and $10^{-13}$ (dotted). The ratios may deviate significantly from unity and have appreciable evolution.}
 \label{fig:truebiratios}
 \end{figure}
 
Now, assuming a predominantly dipolar magnetic field in the NS exterior\,\cite{PaciniNature, GunnOstrikerNature,1969ApJ...157.1395O}, the spin-down due to magnetic-braking torque (from radiation reaction) is given by\,\cite{Shapiro:1983du}
\begin{equation}
I\ns\, \dot{\Omega}\ns = -\frac{1}{6}\, \Omega\ns^3 B^2\ns R\ns^6 \sin^2\alpha \; .
\end{equation}
Here, $I\ns$ is the NS moment-of-inertia and $\alpha$ is the angle between the NS rotation and magnetic axes. Without loss of generality, we take $\alpha=\pi/4$ and neglect the small time dependence that $I\ns$ may have. Specialising to an MG and approximating it to a spinning rigid sphere with $I\mg = \frac{2}{5} M\mg R\mg^2$, we get 
\begin{equation}
\dot{\Omega}\mg(t)=-\frac{5}{24} \frac{R\mg^4}{M\mg} B^2\mg(t) \Omega\mg^3(t) \; .
\vspace{0.05in}
\label{eq:omega}
\end{equation}

We solve the coupled differential equations, Eqs.\,\eqref{eq:bevolv} and \eqref{eq:omega}, for $B\mg(t)$ and $\Omega\mg(t)$ over a time-scale $[1.0,\,1\times10^{5}]\,\mathrm{yrs}$. Based on the solution we may calculate $b_{\text{\tiny{true}}}(t)$ for various $(\Mmcp,\epsilon)$ values. Define the deviation of $b_{\text{\tiny{true}}}(t)$ from the pure magnetic dipole braking index as $\Delta b_{\text{\tiny{true}}}(t) \equiv b_{\text{\tiny{true}}}(t)-3$. The ratio of this quantity, without ($\Delta b^{\cancel{\text{\tiny{SPP}}}}_{\text{\tiny{true}}}$) to that with ($\Delta b^{\text{\tiny{SPP}}}_{\text{\tiny{true}}}$) mCP SPP, is shown in Fig.\,\ref{fig:truebiratios}. The curves are for parametric values $\epsilon=\{10^{-11}, 10^{-12},10^{-13}\}$ and $\Mmcp=10^{-2}\,\mathrm{eV}$. For these values, $F_{\text{\tiny{E}}} > F^{{\text{\tiny{Coul.}}}}_{\text{\tiny{D}}}$. For large $\epsilon$ values, the ratio $\Delta b^{\cancel{\text{\tiny{SPP}}}}_{\text{\tiny{true}}}/\Delta b^{\text{\tiny{SPP}}}_{\text{\tiny{true}}}$ differs significantly from unity and also shows appreciable time evolution.

\section{Summary}{\label{sec:summary}}
Neutron stars may provide unique constraints on exotic particles and interactions. In this paper, we point out a few new ideas to place limits on electrically charged, light,  fermion mCPs by considering their non-perturbative production in neutron stars. As encapsulated in Eqs. (\ref{eq:eloss}) and (\ref{eq:lossoverwhelm}), and Fig. \ref{fig:truebiratios}, the limits are derived based on the condition that the pair production should not drastically change conventional Magnetar physics or evolution. Depending on the mCP parameters, we find that the neutron star energetics, magnetic field evolution and spin-down rates may all be modified. These effects provide a new method whereby light fermion mCPs, below around $1\,\mathrm{eV}$, may be constrained in a relatively model independent way. 

Regarding the affects on Magnetar field decays and spin-down rates, it would be interesting to explore how these may potentially be identified in future observations, and uniquely identified from possible astrophysical processes that may mimic similar signals. It would be also be interesting to investigate these effects further by incorporating a more realistic modelling of the magneto-thermal evolution of the MG system in the presence of mCPs, as well as incorporating any effects of temperature on Schwinger pair-production when $E \shortparallel B$\,\cite{Korwar:2018euc} (in the pure electric or magnetic field case see for instance\,\cite{Brown:2015kgj, Medina:2015qzc, Gould:2017fve, Gould:2017zwi}). Another intriguing avenue to consider is what effects transient electromagnetic fields in the NS magnetosphere may have.
 
\begin{acknowledgments}
We would like to thank Anson Hook and Surjeet Rajendran for discussions. A.T would also like to thank Graham Kribs, Adam Martin, and Tim Cohen for discussions, and thanks the organisers of BBSM-2018 at TIFR, Mumbai for hospitality.
\end{acknowledgments}

\bibliography{JHEP_mcp_SPP_astrophys}

\providecommand{\href}[2]{#2}\begingroup\raggedright\begin{thebibliography}{10}

\bibitem{Goldberg:1986nk}
H.~Goldberg and L.~J. Hall, \emph{{A New Candidate for Dark Matter}},
  \href{https://doi.org/10.1016/0370-2693(86)90731-8}{\emph{Phys. Lett.}
  {\bfseries B174} (1986) 151}.

\bibitem{Cheung:2007ut}
K.~Cheung and T.-C. Yuan, \emph{{Hidden fermion as milli-charged dark matter in
  Stueckelberg Z- prime model}},
  \href{https://doi.org/10.1088/1126-6708/2007/03/120}{\emph{JHEP} {\bfseries
  03} (2007) 120} [\href{https://arxiv.org/abs/hep-ph/0701107}{{\ttfamily
  hep-ph/0701107}}].

\bibitem{Feldman:2007wj}
D.~Feldman, Z.~Liu and P.~Nath, \emph{{The Stueckelberg Z-prime Extension with
  Kinetic Mixing and Milli-Charged Dark Matter From the Hidden Sector}},
  \href{https://doi.org/10.1103/PhysRevD.75.115001}{\emph{Phys. Rev.}
  {\bfseries D75} (2007) 115001}
  [\href{https://arxiv.org/abs/hep-ph/0702123}{{\ttfamily hep-ph/0702123}}].

\bibitem{Holdom:1985ag}
B.~Holdom, \emph{{Two U(1)'s and Epsilon Charge Shifts}},
  \href{https://doi.org/10.1016/0370-2693(86)91377-8}{\emph{Phys. Lett.}
  {\bfseries B166} (1986) 196}.

\bibitem{Dienes:1996zr}
K.~R. Dienes, C.~F. Kolda and J.~March-Russell, \emph{{Kinetic mixing and the
  supersymmetric gauge hierarchy}},
  \href{https://doi.org/10.1016/S0550-3213(97)80028-4,
  10.1016/S0550-3213(97)00173-9}{\emph{Nucl. Phys.} {\bfseries B492} (1997)
  104} [\href{https://arxiv.org/abs/hep-ph/9610479}{{\ttfamily
  hep-ph/9610479}}].

\bibitem{Abel:2003ue}
S.~A. Abel and B.~W. Schofield, \emph{{Brane anti-brane kinetic mixing,
  millicharged particles and SUSY breaking}},
  \href{https://doi.org/10.1016/j.nuclphysb.2004.02.037}{\emph{Nucl. Phys.}
  {\bfseries B685} (2004) 150}
  [\href{https://arxiv.org/abs/hep-th/0311051}{{\ttfamily hep-th/0311051}}].

\bibitem{Batell:2005wa}
B.~Batell and T.~Gherghetta, \emph{{Localized U(1) gauge fields, millicharged
  particles, and holography}},
  \href{https://doi.org/10.1103/PhysRevD.73.045016}{\emph{Phys. Rev.}
  {\bfseries D73} (2006) 045016}
  [\href{https://arxiv.org/abs/hep-ph/0512356}{{\ttfamily hep-ph/0512356}}].

\bibitem{Aldazabal:2000sa}
G.~Aldazabal, L.~E. Ibanez, F.~Quevedo and A.~M. Uranga, \emph{{D-branes at
  singularities: A Bottom up approach to the string embedding of the standard
  model}}, \href{https://doi.org/10.1088/1126-6708/2000/08/002}{\emph{JHEP}
  {\bfseries 08} (2000) 002}
  [\href{https://arxiv.org/abs/hep-th/0005067}{{\ttfamily hep-th/0005067}}].

\bibitem{Abel:2008ai}
S.~A. Abel, M.~D. Goodsell, J.~Jaeckel, V.~V. Khoze and A.~Ringwald,
  \emph{{Kinetic Mixing of the Photon with Hidden U(1)s in String
  Phenomenology}},
  \href{https://doi.org/10.1088/1126-6708/2008/07/124}{\emph{JHEP} {\bfseries
  07} (2008) 124} [\href{https://arxiv.org/abs/0803.1449}{{\ttfamily
  0803.1449}}].

\bibitem{Zavattini:2005tm}
{\scshape PVLAS} collaboration, \emph{{Experimental observation of optical
  rotation generated in vacuum by a magnetic field}},
  \href{https://doi.org/10.1103/PhysRevLett.99.129901,
  10.1103/PhysRevLett.96.110406}{\emph{Phys. Rev. Lett.} {\bfseries 96} (2006)
  110406} [\href{https://arxiv.org/abs/hep-ex/0507107}{{\ttfamily
  hep-ex/0507107}}].

\bibitem{Adriani:2008zr}
{\scshape PAMELA} collaboration, \emph{{An anomalous positron abundance in
  cosmic rays with energies 1.5-100 GeV}},
  \href{https://doi.org/10.1038/nature07942}{\emph{Nature} {\bfseries 458}
  (2009) 607} [\href{https://arxiv.org/abs/0810.4995}{{\ttfamily 0810.4995}}].

\bibitem{Chang:2008aa}
J.~Chang et~al., \emph{{An excess of cosmic ray electrons at energies of
  300-800 GeV}}, \href{https://doi.org/10.1038/nature07477}{\emph{Nature}
  {\bfseries 456} (2008) 362}.

\bibitem{Bowman:2018yin}
J.~D. Bowman, A.~E.~E. Rogers, R.~A. Monsalve, T.~J. Mozdzen and N.~Mahesh,
  \emph{{An absorption profile centred at 78 megahertz in the sky-averaged
  spectrum}}, \href{https://doi.org/10.1038/nature25792}{\emph{Nature}
  {\bfseries 555} (2018) 67}.

\bibitem{Barkana:2018lgd}
R.~Barkana, \emph{{Possible interaction between baryons and dark-matter
  particles revealed by the first stars}},
  \href{https://doi.org/10.1038/nature25791}{\emph{Nature} {\bfseries 555}
  (2018) 71} [\href{https://arxiv.org/abs/1803.06698}{{\ttfamily 1803.06698}}].

\bibitem{Jaeckel:2010ni}
J.~Jaeckel and A.~Ringwald, \emph{{The Low-Energy Frontier of Particle
  Physics}},
  \href{https://doi.org/10.1146/annurev.nucl.012809.104433}{\emph{Ann. Rev.
  Nucl. Part. Sci.} {\bfseries 60} (2010) 405}
  [\href{https://arxiv.org/abs/1002.0329}{{\ttfamily 1002.0329}}].

\bibitem{Collar:2012olx}
J.~I. Collar et~al., \emph{{New light, weakly-coupled particles}},  in
  \emph{{Fundamental Physics at the Intensity Frontier}}.

\bibitem{Hook:2017vyc}
A.~Hook and J.~Huang, \emph{{Bounding millimagnetically charged particles with
  magnetars}}, \href{https://doi.org/10.1103/PhysRevD.96.055010}{\emph{Phys.
  Rev.} {\bfseries D96} (2017) 055010}
  [\href{https://arxiv.org/abs/1705.01107}{{\ttfamily 1705.01107}}].

\bibitem{Gould:2017zwi}
O.~Gould and A.~Rajantie, \emph{{Magnetic monopole mass bounds from heavy ion
  collisions and neutron stars}},
  \href{https://doi.org/10.1103/PhysRevLett.119.241601}{\emph{Phys. Rev. Lett.}
  {\bfseries 119} (2017) 241601}
  [\href{https://arxiv.org/abs/1705.07052}{{\ttfamily 1705.07052}}].

\bibitem{Gies:2006hv}
H.~Gies, J.~Jaeckel and A.~Ringwald, \emph{{Accelerator Cavities as a Probe of
  Millicharged Particles}},
  \href{https://doi.org/10.1209/epl/i2006-10356-5}{\emph{Europhys. Lett.}
  {\bfseries 76} (2006) 794}
  [\href{https://arxiv.org/abs/hep-ph/0608238}{{\ttfamily hep-ph/0608238}}].

\bibitem{Davidson:1993sj}
S.~Davidson and M.~E. Peskin, \emph{{Astrophysical bounds on millicharged
  particles in models with a paraphoton}},
  \href{https://doi.org/10.1103/PhysRevD.49.2114}{\emph{Phys. Rev.} {\bfseries
  D49} (1994) 2114} [\href{https://arxiv.org/abs/hep-ph/9310288}{{\ttfamily
  hep-ph/9310288}}].

\bibitem{Davidson:2000hf}
S.~Davidson, S.~Hannestad and G.~Raffelt, \emph{{Updated bounds on millicharged
  particles}}, \href{https://doi.org/10.1088/1126-6708/2000/05/003}{\emph{JHEP}
  {\bfseries 05} (2000) 003}
  [\href{https://arxiv.org/abs/hep-ph/0001179}{{\ttfamily hep-ph/0001179}}].

\bibitem{Vogel:2013raa}
H.~Vogel and J.~Redondo, \emph{{Dark Radiation constraints on minicharged
  particles in models with a hidden photon}},
  \href{https://doi.org/10.1088/1475-7516/2014/02/029}{\emph{JCAP} {\bfseries
  1402} (2014) 029} [\href{https://arxiv.org/abs/1311.2600}{{\ttfamily
  1311.2600}}].

\bibitem{Foot:2014uba}
R.~Foot and S.~Vagnozzi, \emph{{Dissipative hidden sector dark matter}},
  \href{https://doi.org/10.1103/PhysRevD.91.023512}{\emph{Phys. Rev.}
  {\bfseries D91} (2015) 023512}
  [\href{https://arxiv.org/abs/1409.7174}{{\ttfamily 1409.7174}}].

\bibitem{Masso:2006gc}
E.~Masso and J.~Redondo, \emph{{Compatibility of CAST search with axion-like
  interpretation of PVLAS results}},
  \href{https://doi.org/10.1103/PhysRevLett.97.151802}{\emph{Phys. Rev. Lett.}
  {\bfseries 97} (2006) 151802}
  [\href{https://arxiv.org/abs/hep-ph/0606163}{{\ttfamily hep-ph/0606163}}].

\bibitem{Abel:2006qt}
S.~A. Abel, J.~Jaeckel, V.~V. Khoze and A.~Ringwald, \emph{{Illuminating the
  Hidden Sector of String Theory by Shining Light through a Magnetic Field}},
  \href{https://doi.org/10.1016/j.physletb.2008.03.076}{\emph{Phys. Lett.}
  {\bfseries B666} (2008) 66}
  [\href{https://arxiv.org/abs/hep-ph/0608248}{{\ttfamily hep-ph/0608248}}].

\bibitem{Foot:2007cq}
R.~Foot and A.~Kobakhidze, \emph{{A Simple explanation of the PVLAS anomaly in
  spontaneously broken mirror models}},
  \href{https://doi.org/10.1016/j.physletb.2007.04.045}{\emph{Phys. Lett.}
  {\bfseries B650} (2007) 46}
  [\href{https://arxiv.org/abs/hep-ph/0702125}{{\ttfamily hep-ph/0702125}}].

\bibitem{Melchiorri:2007sq}
A.~Melchiorri, A.~Polosa and A.~Strumia, \emph{{New bounds on millicharged
  particles from cosmology}},
  \href{https://doi.org/10.1016/j.physletb.2007.05.042}{\emph{Phys. Lett.}
  {\bfseries B650} (2007) 416}
  [\href{https://arxiv.org/abs/hep-ph/0703144}{{\ttfamily hep-ph/0703144}}].

\bibitem{1934PNAS...20..254B}
W.~{Baade} and F.~{Zwicky}, \emph{{On Super-novae}},
  \href{https://doi.org/10.1073/pnas.20.5.254}{\emph{Proceedings of the
  National Academy of Science} {\bfseries 20} (1934) 254}.

\bibitem{1934PNAS...20..259B}
W.~{Baade} and F.~{Zwicky}, \emph{{Cosmic Rays from Super-novae}},
  \href{https://doi.org/10.1073/pnas.20.5.259}{\emph{Proceedings of the
  National Academy of Science} {\bfseries 20} (1934) 259}.

\bibitem{Becker:2009}
W.~{Becker}~(ed.), \emph{{Neutron Stars and Pulsars}}. Springer, Berlin,
  Germany, 2009.

\bibitem{Mereghetti:2008je}
S.~Mereghetti, \emph{{The strongest cosmic magnets: Soft Gamma-ray Repeaters
  and Anomalous X-ray Pulsars}},
  \href{https://doi.org/10.1007/s00159-008-0011-z}{\emph{Astron. Astrophys.
  Rev.} {\bfseries 15} (2008) 225}
  [\href{https://arxiv.org/abs/0804.0250}{{\ttfamily 0804.0250}}].

\bibitem{1992ApJ...392L...9D}
R.~C. {Duncan} and C.~{Thompson}, \emph{{Formation of very strongly magnetized
  neutron stars - Implications for gamma-ray bursts}},
  \href{https://doi.org/10.1086/186413}{\emph{\apj} {\bfseries 392} (1992) L9}.

\bibitem{1993ApJ...408..194T}
C.~{Thompson} and R.~C. {Duncan}, \emph{{Neutron star dynamos and the origins
  of pulsar magnetism}}, \href{https://doi.org/10.1086/172580}{\emph{\apj}
  {\bfseries 408} (1993) 194}.

\bibitem{Thompson:1995gw}
C.~Thompson and R.~C. Duncan, \emph{{The Soft gamma repeaters as very strongly
  magnetized neutron stars - 1. Radiative mechanism for outbursts}},
  {\emph{Mon. Not. Roy. Astron. Soc.} {\bfseries 275} (1995) 255}.

\bibitem{2004hpa..book.....L}
D.~R. {Lorimer} and M.~{Kramer}, \emph{{Handbook of Pulsar Astronomy}}.
  Cambridge University Press, Cambridge, UK, Dec., 2004.

\bibitem{Goldreich:1969sb}
P.~Goldreich and W.~H. Julian, \emph{{Pulsar electrodynamics}},
  \href{https://doi.org/10.1086/150119}{\emph{Astrophys. J.} {\bfseries 157}
  (1969) 869}.

\bibitem{Harding:1998ma}
A.~K. Harding and A.~G. Muslimov, \emph{{Particle acceleration zones above
  pulsar polar caps: electron and positron pair formation fronts}},
  \href{https://doi.org/10.1086/306394}{\emph{Astrophys. J.} {\bfseries 508}
  (1998) 328} [\href{https://arxiv.org/abs/astro-ph/9805132}{{\ttfamily
  astro-ph/9805132}}].

\bibitem{Dyks:2000ee}
J.~Dyks and B.~Rudak, \emph{{Approximate expressions for polar gap electric
  field of pulsars}}, {\emph{Astron. Astrophys.} {\bfseries 362} (2000) 1004}
  [\href{https://arxiv.org/abs/astro-ph/0006256}{{\ttfamily
  astro-ph/0006256}}].

\bibitem{Ruderman:1975ju}
M.~A. Ruderman and P.~G. Sutherland, \emph{{Theory of pulsars: Polar caps,
  sparks, and coherent microwave radiation}},
  \href{https://doi.org/10.1086/153393}{\emph{Astrophys. J.} {\bfseries 196}
  (1975) 51}.

\bibitem{Hibschman:2001kn}
J.~A. Hibschman and J.~Arons, \emph{{Pair multiplicities and pulsar death}},
  \href{https://doi.org/10.1086/321378}{\emph{Astrophys. J.} {\bfseries 554}
  (2001) 624} [\href{https://arxiv.org/abs/astro-ph/0102175}{{\ttfamily
  astro-ph/0102175}}].

\bibitem{Nikishov:1970br}
A.~I. Nikishov, \emph{{Barrier scattering in field theory removal of klein
  paradox}}, \href{https://doi.org/10.1016/0550-3213(70)90527-4}{\emph{Nucl.
  Phys.} {\bfseries B21} (1970) 346}.

\bibitem{Kim:2003qp}
S.~P. Kim and D.~N. Page, \emph{{Schwinger pair production in electric and
  magnetic fields}},
  \href{https://doi.org/10.1103/PhysRevD.73.065020}{\emph{Phys. Rev.}
  {\bfseries D73} (2006) 065020}
  [\href{https://arxiv.org/abs/hep-th/0301132}{{\ttfamily hep-th/0301132}}].

\bibitem{Dunne:2004nc}
G.~V. Dunne, \emph{{Heisenberg-Euler effective Lagrangians: Basics and
  extensions}},  in \emph{From fields to strings: Circumnavigating theoretical
  physics. Ian Kogan memorial collection (3 volume set)}, M.~Shifman,
  A.~Vainshtein and J.~Wheater, eds., pp.~445--522, (2004),
  \href{https://arxiv.org/abs/hep-th/0406216}{{\ttfamily hep-th/0406216}},
  \href{https://doi.org/10.1142/9789812775344_0014}{DOI}.

\bibitem{Ruffini:2009hg}
R.~Ruffini, G.~Vereshchagin and S.-S. Xue, \emph{{Electron-positron pairs in
  physics and astrophysics: from heavy nuclei to black holes}},
  \href{https://doi.org/10.1016/j.physrep.2009.10.004}{\emph{Phys. Rept.}
  {\bfseries 487} (2010) 1} [\href{https://arxiv.org/abs/0910.0974}{{\ttfamily
  0910.0974}}].

\bibitem{Korwar:2018euc}
M.~Korwar and A.~M. Thalapillil, \emph{{Finite temperature Schwinger pair
  production in coexistent electric and magnetic fields}},
  \href{https://doi.org/10.1103/PhysRevD.98.076016}{\emph{Phys. Rev.}
  {\bfseries D98} (2018) 076016}
  [\href{https://arxiv.org/abs/1808.01295}{{\ttfamily 1808.01295}}].

\bibitem{Schwinger:1951nm}
J.~S. Schwinger, \emph{{On gauge invariance and vacuum polarization}},
  \href{https://doi.org/10.1103/PhysRev.82.664}{\emph{Phys. Rev.} {\bfseries
  82} (1951) 664}.

\bibitem{Olausen:2013bpa}
S.~A. Olausen and V.~M. Kaspi, \emph{{The McGill Magnetar Catalog}},
  \href{https://doi.org/10.1088/0067-0049/212/1/6}{\emph{Astrophys. J. Suppl.}
  {\bfseries 212} (2014) 6} [\href{https://arxiv.org/abs/1309.4167}{{\ttfamily
  1309.4167}}].

\bibitem{Kaspi:2017fwg}
V.~M. Kaspi and A.~Beloborodov, \emph{{Magnetars}},
  \href{https://doi.org/10.1146/annurev-astro-081915-023329}{\emph{Ann. Rev.
  Astron. Astrophys.} {\bfseries 55} (2017) 261}
  [\href{https://arxiv.org/abs/1703.00068}{{\ttfamily 1703.00068}}].

\bibitem{Affleck:1981bma}
I.~K. Affleck, O.~Alvarez and N.~S. Manton, \emph{{Pair Production at Strong
  Coupling in Weak External Fields}},
  \href{https://doi.org/10.1016/0550-3213(82)90455-2}{\emph{Nucl. Phys.}
  {\bfseries B197} (1982) 509}.

\bibitem{Affleck:1981ag}
I.~K. Affleck and N.~S. Manton, \emph{{Monopole Pair Production in a Magnetic
  Field}}, \href{https://doi.org/10.1016/0550-3213(82)90511-9}{\emph{Nucl.
  Phys.} {\bfseries B194} (1982) 38}.

\bibitem{Ritus:1998jm}
V.~I. Ritus, \emph{{Effective Lagrange function of intense electromagnetic
  field in QED}},  in \emph{{Frontier tests of QED and physics of the vacuum.
  Proceedings, Workshop, Sandansky, Bulgaria, June 9-15, 1998}}.

\bibitem{Li:2013pfh}
X.~Li and M.~B. Voloshin, \emph{{Electric discharge in vacuum by minicharged
  particles}}, \href{https://doi.org/10.1142/S0217732314500540}{\emph{Mod.
  Phys. Lett.} {\bfseries A29} (2014) 1450054}
  [\href{https://arxiv.org/abs/1401.0049}{{\ttfamily 1401.0049}}].

\bibitem{Shapiro:1983du}
S.~L. Shapiro and S.~A. Teukolsky, \emph{{Black holes, white dwarfs, and
  neutron stars: The physics of compact objects}}. 1983.

\bibitem{1992ApJ...395..250G}
P.~{Goldreich} and A.~{Reisenegger}, \emph{{Magnetic field decay in isolated
  neutron stars}}, \href{https://doi.org/10.1086/171646}{\emph{\apj} {\bfseries
  395} (1992) 250}.

\bibitem{Glampedakis:2010ec}
K.~Glampedakis, D.~I. Jones and L.~Samuelsson, \emph{{Ambipolar diffusion in
  superfluid neutron stars}},
  \href{https://doi.org/10.1111/j.1365-2966.2011.18278.x}{\emph{Mon. Not. Roy.
  Astron. Soc.} {\bfseries 413} (2011) 2021}
  [\href{https://arxiv.org/abs/1010.1153}{{\ttfamily 1010.1153}}].

\bibitem{Turolla:2015mwa}
R.~Turolla, S.~Zane and A.~Watts, \emph{{Magnetars: the physics behind
  observations. A review}},
  \href{https://doi.org/10.1088/0034-4885/78/11/116901}{\emph{Rept. Prog.
  Phys.} {\bfseries 78} (2015) 116901}
  [\href{https://arxiv.org/abs/1507.02924}{{\ttfamily 1507.02924}}].

\bibitem{Aguilera:2007xk}
D.~N. Aguilera, J.~A. Pons and J.~A. Miralles, \emph{{2D Cooling of Magnetized
  Neutron Stars}},
  \href{https://doi.org/10.1051/0004-6361:20078786}{\emph{Astron. Astrophys.}
  {\bfseries 486} (2008) 255}
  [\href{https://arxiv.org/abs/0710.0854}{{\ttfamily 0710.0854}}].

\bibitem{Vigano:2013lea}
D.~Vigano, N.~Rea, J.~A. Pons, R.~Perna, D.~N. Aguilera and J.~A. Miralles,
  \emph{{Unifying the observational diversity of isolated neutron stars via
  magneto-thermal evolution models}},
  \href{https://doi.org/10.1093/mnras/stt1008}{\emph{Mon. Not. Roy. Astron.
  Soc.} {\bfseries 434} (2013) 123}
  [\href{https://arxiv.org/abs/1306.2156}{{\ttfamily 1306.2156}}].

\bibitem{2007A&A...470..303P}
J.~A. {Pons} and U.~{Geppert}, \emph{{Magnetic field dissipation in neutron
  star crusts: from magnetars to isolated neutron stars}},
  \href{https://doi.org/10.1051/0004-6361:20077456}{\emph{Astronomy and
  Astrophysics} {\bfseries 470} (2007) 303}
  [\href{https://arxiv.org/abs/astro-ph/0703267}{{\ttfamily
  astro-ph/0703267}}].

\bibitem{PaciniNature}
F.~{Pacini}, \emph{{Energy Emission from a Neutron Star}},
  \href{https://doi.org/10.1038/216567a0}{\emph{Nature} {\bfseries 216} (1967)
  567}.

\bibitem{GunnOstrikerNature}
J.~P. {Ostriker} and J.~E. {Gunn}, \emph{{Magnetic Dipole Radiation from
  Pulsars}}, \href{https://doi.org/10.1038/221454a0}{\emph{Nature} {\bfseries
  221} (1969) 454}.

\bibitem{1969ApJ...157.1395O}
J.~P. {Ostriker} and J.~E. {Gunn}, \emph{{On the Nature of Pulsars. I.
  Theory}}, \href{https://doi.org/10.1086/150160}{\emph{\apj} {\bfseries 157}
  (1969) 1395}.

\bibitem{Brown:2015kgj}
A.~R. Brown, \emph{{Schwinger pair production at nonzero temperatures or in
  compact directions}},
  \href{https://doi.org/10.1103/PhysRevD.98.036008}{\emph{Phys. Rev.}
  {\bfseries D98} (2018) 036008}
  [\href{https://arxiv.org/abs/1512.05716}{{\ttfamily 1512.05716}}].

\bibitem{Medina:2015qzc}
L.~Medina and M.~C. Ogilvie, \emph{{Schwinger Pair Production at Finite
  Temperature}}, \href{https://doi.org/10.1103/PhysRevD.95.056006}{\emph{Phys.
  Rev.} {\bfseries D95} (2017) 056006}
  [\href{https://arxiv.org/abs/1511.09459}{{\ttfamily 1511.09459}}].

\bibitem{Gould:2017fve}
O.~Gould and A.~Rajantie, \emph{{Thermal Schwinger pair production at arbitrary
  coupling}}, \href{https://doi.org/10.1103/PhysRevD.96.076002}{\emph{Phys.
  Rev.} {\bfseries D96} (2017) 076002}
  [\href{https://arxiv.org/abs/1704.04801}{{\ttfamily 1704.04801}}].

\end{thebibliography}\endgroup
\bibliographystyle{JHEP}

\end{document}